\documentclass{PoS}

\newcommand{\insertfig}[3]{
\begin{figure}
\begin{center}
\includegraphics[angle=0,width=0.7\textwidth]{#1}
\caption{#2}
\label{#3}
\end{center}
\end{figure}
}

\title{Charmonium spectral functions in $N_f=2$ QCD at high temperature}

\ShortTitle{Charmonium spectral functions in $N_f=2$ QCD at high temperature}

\author{G.~Aarts, C.R.~Allton\thanks{Speaker, email: c.allton@swansea.ac.uk} \\
        Department of Physics\\
	Swansea University\\
	Swansea, U.K.}
\author{R.~Morrin, A.P.~\'O Cais, M.B.~Oktay, M.J.~Peardon, J.I.~Skullerud\\
        School of Mathematics\\
	Trinity College\\
	Dublin, Ireland}

\abstract{Charmonium systems in two-flavour QCD at non-zero
temperature are studied at, and above the deconfining
transition. Using anisotropic lattices the MEM approach is used to
extract spectral functions for these channels.  By carefully varying
some of the irrelevant parameters in the MEM procedure, we confirm
that our systematic effects are under control.  The $\eta_c$ and
$J/\psi$ states are found to persist at temperatures $\gtrsim 1.3
T_c$ in agreement with our previous dynamical studies.}

\FullConference{XXIVth International Symposium on Lattice Field Theory\\
                July 23-28, 2006\\
                Tucson, Arizona, USA}

\begin{document}



\section{Introduction}

The recent experimental evidence of the quark-gluon plasma phase of
QCD in RHIC \cite{RHIC} provides experimentalists
and theorists with a formidable challenge: How is the physics of this
new phase uncovered from the products of the fireball when these
products have necessarily condensed back into the confined phase?

To answer this question it is crucial to know whether hadronic
states in fact persist in the quark-gluon plasma phase. Current understanding
of the plasma phase suggests that while it is not confining,
it is still strongly interacting, suggesting that hadronic states can
indeed survive above the de-confining temperature, $T_c$ \cite{scqgp}.

This study investigates mesonic states at and above $T_c$ via their
spectral functions, $\rho(\omega,\mathbf{p})$. $\rho$ can be defined from
Euclidean hadronic 2--point function, $G(t,\mathbf{p})$, as
\begin{equation}
G(t,\mathbf{p}) = \int_0^\infty \rho(\omega,\mathbf{p})\; K(t,\omega)\;
\frac{d\omega}{2\pi}
\end{equation}
where the kernel, $K(t,\omega)$, at temperature, $T$, is given by
\[
K(t,\omega) = \frac{\cosh[\omega(t-1/(2T))]}{\sinh[\omega/(2T)]}.
\]
Spectral functions contain information on the stability of hadronic
states, transport coefficients and dilepton production rates, and thus
are fundamental physical quantities of the system.  They can be
derived from lattice calculations of hadronic correlators via the
Maximum Entropy Method (MEM)\cite{mem}. However, the success of this
approach relies on the quality of the input data, in particular on
having estimates of $G(t)$ over a large time range. Using isotropic
lattices, this requirement runs at odds with finite temperature
studies which restrict the temporal extent to be small.  We overcome
this conflict by using {\em anisotropic} lattices, to ensure that the
number of time points, $N_t$, is large, while maintaining a non-zero
temperature, $T = 1/(a_t N_t)$ (where $a_t$ is the temporal lattice
spacing which we take much smaller than the spatial spacing, $a_s$).
Using this approach, we have performed dynamical studies of charmonium
states at temperatures $T_c \lesssim T \lesssim 2T_c$ and are able to
confirm that they are still bound even for temperatures above
$T_c$. This work is a continuation of that presented in
\cite{us}.



\section{Lattice Parameters}

Because our lattices used dynamical fermions, tuning the two anisotropy
parameters in the gauge and quark actions is highly non-trivial. This is
because there is feedback from the fermions to the gluons which is not
present in the quenched anisotropic case. Details of this procedure for
the choice of light quark mass and gauge coupling used in this work can
be found in \cite{dublin}.
The ensembles studied are more highly tuned than in our previous work
presented in \cite{us} and have a renormalised anisotropy
for both quarks and gluons very close to 6.

A highly improved anisotropic gauge action was used with
Wilson~+~Hamber-Wu fermions with stout links. The details of this
action can be found in \cite{dublin}. Table \ref{tab:params} lists the
parameters used in the simulations.  As can be seen, three
temperatures were studied: $T/T_c \sim 1,\; 1.3,\; 2$ and two spatial
volumes were simulated enabling a finite volume analysis.  Two heavy
quark masses were studied, with the $am_c=0.080$ being closer to the
experimental charm quark.


\begin{table}
\begin{center}
\begin{tabular}{lcl}
\hline
Light quarks              & $M_\pi/M_\rho$& $\sim 0.5$ \\
Heavy quarks              & $am_c$        & $0.080, 0.092$ \\
(Renormalised) Anisotropy & $\xi$         & $6$ \\
Lattice spacings$\;\;$    & $a_t$         & $\sim 0.025$ fm \\
                          & $a_s$         & $\sim 0.15$ fm \\
Spatial Volume            & $N_s^3$       & $8^3 \;\; (\& 12^3)$ \\
Temporal Extent           & $N_t$         & $16$ \hspace{5mm} $\longleftrightarrow$ \hspace{5mm} $\sim \;2\;T_c$ \\
                          &               & $24$ \hspace{5mm} $\longleftrightarrow$ \hspace{5mm} $\sim \;1.3\;T_c$ \\
                          &               & $32$ \hspace{5mm} $\longleftrightarrow$ \hspace{5mm} $\sim \;T_c$ \\
Statistics                & $N_{cfg}$     & $\sim 500$ \\
\hline
\end{tabular}
\end{center}
\caption{The lattice parameters used.}
\label{tab:params}
\end{table}


The deconfining transition was found between $N_t = 34$ and $33$ by
studying the Polyakov line on the $N_s^3 = 12^3$ spatial volume.  (No
discontinuity in the Polyakov line was found in the $N_s^3 = 8^3$ case.)

Before embarking on the full simulation, it is worth studying the
systematic effects induced by the lattice discretisation.  Figure
\ref{fig:free} shows the pseudoscalar and vector spectral functions
for the free theory for both the continuum and lattice. (See
\cite{gert} for details of how this calculation was performed.) In the
lattice case, the anisotropy of 6 was chosen and two temporal extents,
$N_t = 24$ and $32$ are shown. As can be seen, the lattice functions
reproduce the continuum ones for small and moderate energies, but
there is an unphysical cusp at $a_t\omega \sim 0.7$ which is
independent of $N_t$ and channel. This leads us to conclude that
lattice spectral functions above this $a_t\omega$ value are likely to
be contaminated by systematic effects. Note also that the cusp for the
run parameters in \cite{us} occurs at the smaller value of $a_t\omega
\sim 0.6$, and hence the results presented here are closer to the
continuum ones over a slightly larger energy range.


\insertfig{TCDxi6.eps}
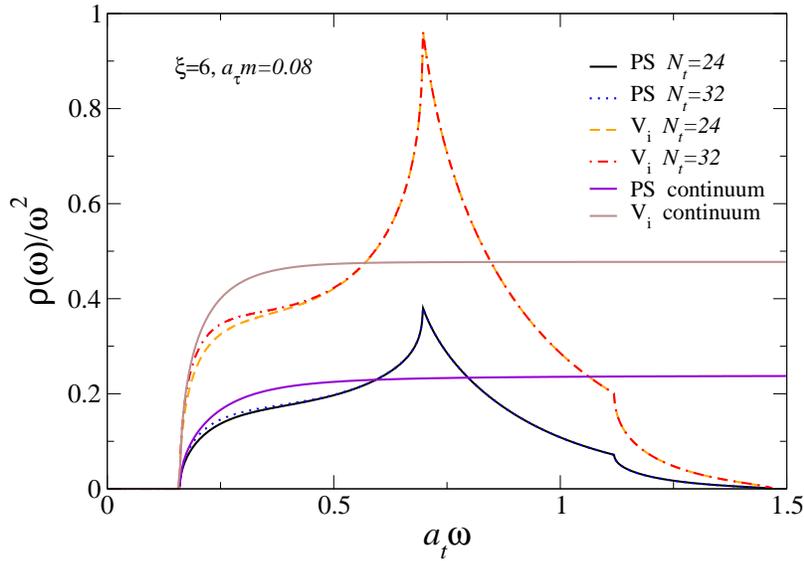
{The spectral function for the vector and pseudoscalar mesons
in free field theory for the parameter values in this work.
The continuum results and those
from the lattice with $N_t = 24$ and $32$ are shown.}
{fig:free}




\section{Results}

Since we have correlation functions calculated for a number of different
parameters, we can study how the spectral functions vary with:

\begin{itemize}

\item energy resolution for the MEM

\item starting time for the MEM analysis

\item spatial volume

\item heavy quark mass

\item channel (i.e. pseudoscalar, vector, axial, scalar)

\item temperature

\end{itemize}

\begin{figure}
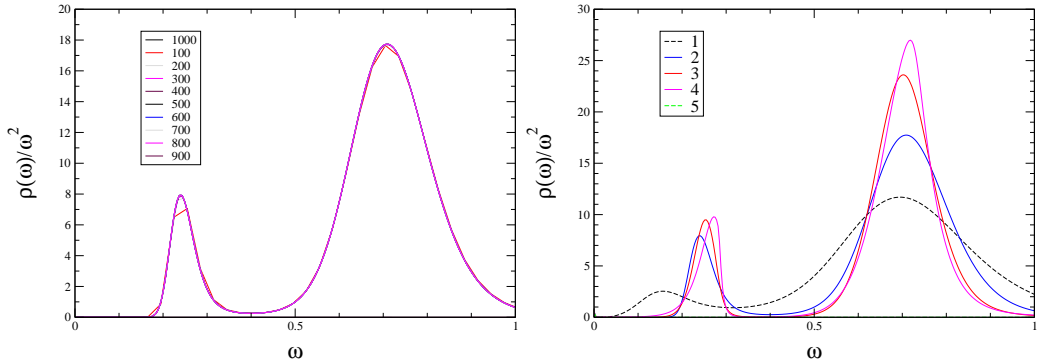

\begin{center}
\begin{minipage}[t]{0.45\textwidth}
\begin{center}
\includegraphics[width=\textwidth]{m092_D8x24_sc_vary_nw_2-11.eps}
\end{center}
\end{minipage}
\begin{minipage}[t]{0.45\textwidth}
\begin{center}
\includegraphics[width=\textwidth]{m092_D8x24_sc_900_vary_t1-11.eps}
\end{center}
\end{minipage}
\end{center}
\vspace{-5mm}
\caption{
{\bf LEFT:}
The spectral function with various energy resolutions, i.e. the number
of discretised $\omega$ points, $N_\omega$, showing the insensitivity
of the spectral function to this quantity. The scalar channel with
$am_c=0.092$ on the $8^3 \times 24$ lattice was used.
{\bf RIGHT:}
The spectral function with various starting times, $t_{\mbox{start}}$
for the MEM analysis.  This shows the insensitivity of the spectral
function to $t_{\mbox{start}}$ within the range
$2 \le t_{\mbox{start}} \le 4$.
(The same channel as the left panel was used.)
}
\label{fig:vary_nw_t1}
\end{figure}

The hope is that the spectral function determined from MEM will {\em
not} vary with the first three items since they are unphysical, and
this is what we find. The left panel in figure \ref{fig:vary_nw_t1}
shows the total lack of effect when the energy resolution of the MEM
is varied from $n_\omega = 100$ through to 1000.


In the right panel of figure \ref{fig:vary_nw_t1} the effect of
varying the start time, $t_{\mbox{start}}$, of the MEM is
studied. Note that the spectral functions for $t_{\mbox{start}} = 2,
3$ or 4 are all compatible, i.e. there is a range of sensible
$t_{\mbox{start}}$ values which produce consistent results. However,
for $t_{\mbox{start}}=1$ and $t_{\mbox{start}} \ge 5$ the spectral
functions do vary.  This can be understood since there are contact
terms which enter the $t_{\mbox{start}}=1$ case, and if
$t_{\mbox{start}} \ge 5$ is used, there is little signal left for the
MEM to latch onto ($N_t = 24$ in this case).

\begin{figure}
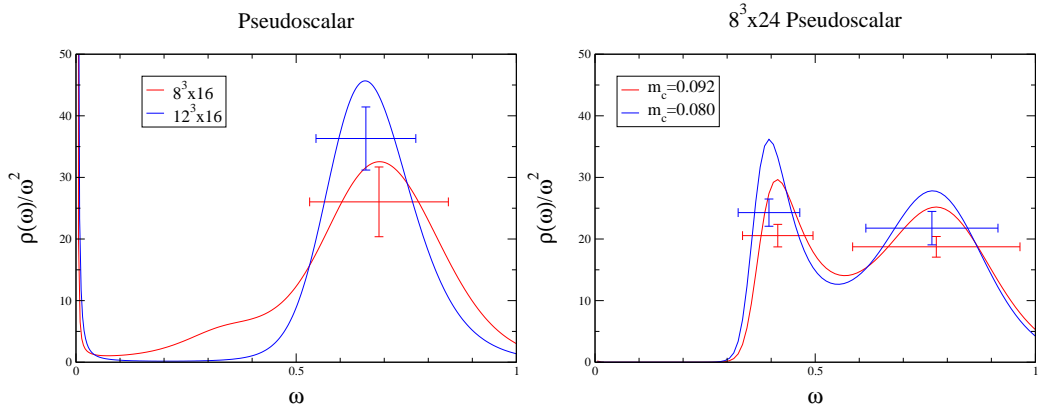

\begin{center}
\begin{minipage}[t]{0.45\textwidth}
\begin{center}
\includegraphics[width=\textwidth]{m092_vary_vol_p5_2-7.eps}
\end{center}
\end{minipage}
\begin{minipage}[t]{0.45\textwidth}
\begin{center}
\includegraphics[width=\textwidth]{vary_m_D8x24_p5.eps}
\end{center}
\end{minipage}
\end{center}
\vspace{-5mm}
\caption{
{\bf LEFT:}
The spectral function for both spatial volumes $N_s = 8$ and
$12$ illustrating the small finite volume effects. The
pseudoscalar channel with $N_t=16$ is shown.
{\bf RIGHT:}
The spectral function with both $am_c = 0.08$ and 0.092
showing the small sensitivity to the $m_c$ value. The
pseudoscalar channel on the $8^3 \times 24$ lattice was used.}
\label{fig:vary_ns_mc}
\end{figure}

Finite volume effects are studied in left panel of figure
\ref{fig:vary_ns_mc}.  As can be seen, there is little variation in
the spectral function as $N_s$ is changed, confirming that our results
are not dominated by finite volume systematics.


Now that these lattice systematic effects are studied and have been
shown not to contaminate our spectral functions, we turn to varying
the physical parameters.  The right panel of figure
\ref{fig:vary_ns_mc} shows the effect of varying the charm quark mass
from $am_c = 0.08$ to 0.092. The ground state peak is slightly higher
in the latter case as expected, but otherwise the spectral function
shows little dependency on $m_c$.

\begin{figure}
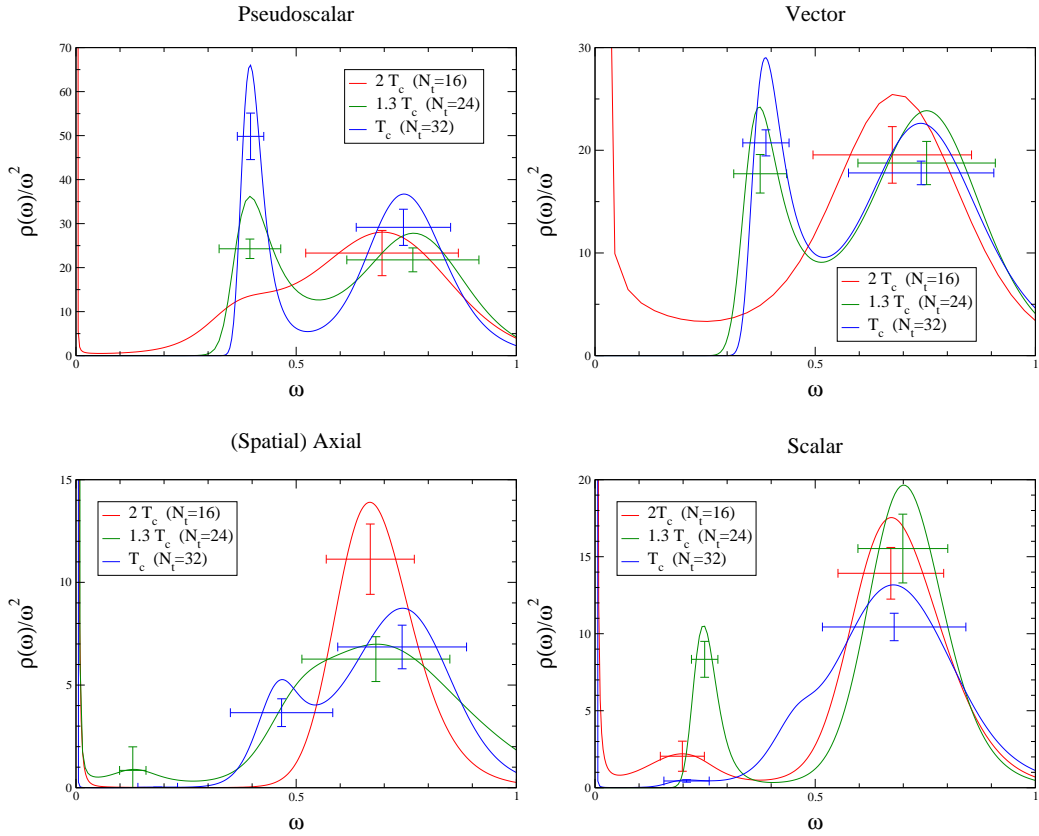

\begin{center}
\begin{minipage}[t]{0.45\textwidth}
\begin{center}
\includegraphics[width=\textwidth]{m080_D8x_varyT_p5.eps}
\end{center}
\end{minipage}
\begin{minipage}[t]{0.45\textwidth}
\begin{center}
\includegraphics[width=\textwidth]{m080_D8x_varyT_vi.eps}
\end{center}
\end{minipage}

\bigskip

\begin{minipage}[t]{0.45\textwidth}
\begin{center}
\includegraphics[width=\textwidth]{m080_D8x_varyT_ai.eps}
\end{center}
\end{minipage}
\begin{minipage}[t]{0.45\textwidth}
\begin{center}
\includegraphics[width=\textwidth]{m080_D8x_varyT_sc.eps}
\end{center}
\end{minipage}
\end{center}
\vspace{-5mm}
\caption{
The meson spectral function at various
temperatures (with $am_c=0.08$ and $N_s = 8$).
}
\label{fig:meson}
\end{figure}

Finally we consider varying the temperature. Figures \ref{fig:meson}
show the spectral functions for the three
temperatures studied for the four mesonic channels. ($N_s=8$
is used throughout.)
As can be seen there is clear evidence of a bound state in the
pseudoscalar and vector channels which survives above $T_c$ and
melts between $1.3\,T_c$ and $T_c$. At $2\,T_c$ there is evidence
of non-zero spectral weight at small energies.
In the case of the axial channel, the state seems to melt by
around $\sim 1.3\, T_c$, and for the scalar channel the situation
is less clear.


\section{Conclusions}

This work continues that first presented in \cite{us}
where anisotropic dynamical lattices were used together with the MEM
technique to determine charmonium spectral functions at non-zero
temperatures.
Improvements over the work described in \cite{us} were
that the anisotropy parameters are now better tuned, a larger
volume considered with more than one heavy quark mass simulated.

We have confirmed the spectral functions from MEM to check that they
are stable for sensible variations in the MEM parameters (energy
discretisation, starting value of the time interval, volume).

Our results confirm our earlier work \cite{us}
and previous quenched studies \cite{quenched}: the pseudoscalar ($\eta_c$)
and vector ($J/\psi$) charmonium states melt between $1.3\,T_c$
and $2\,T_c$, while the scalar ($\chi_{c0}$) and axialvector
($\chi_{c1}$) states are less less certain, but seem to melt by $1.3\, T_c$.

Future work will include varying the default model, inclusion of non-zero
momenta and studies of light hadrons.




\end{document}